# Kaon Condensation and the Non-Uniform Nuclear Matter

Toshiki Maruyama[*], Toshitaka Tatsumi[†], Dmitri N. Voskresensky[**], Tomonori Tanigawa[‡*] and Satoshi Chiba[*]

[*]*Advanced Science Research Center, Japan Atomic Energy Research Institute, Tokai, Ibaraki 319-1195, Japan*
[†]*Department of Physics, Kyoto University, Kyoto 606-8502, Japan*
[**]*Moscow Institute for Physics and Engineering, Kashirskoe sh. 31, Moscow 115409, Russia*
[‡]*Japan Society for the Promotion of Science, Tokyo 102-8471, Japan*

**Abstract.**
Non-uniform structures of nuclear matter are studied in a wide density-range. Using the density functional theory with a relativistic mean-field model, we examine non-uniform structures at sub-nuclear densities (nuclear "pastas") and at high densities, where kaon condensate is expected. We try to give a unified view about the change of the matter structure as density increases, carefully taking into account the Coulomb screening effects from the viewpoint of first-order phase transition.

## INTRODUCTION

There have been discussed various phase transitions in nuclear matter, like liquid-gas or neutron-drip phase transition, meson condensations, hadron-quark deconfinement transition, etc. In most cases they exhibit the first-order phase transitions. In the first-order phase transitions with more than one chemical potential, the structured mixed phase may be expected by way of the Gibbs conditions for the phase equilibrium [1].

At sub-nuclear densities, exotic nuclear shapes, called nuclear "pastas", are expected: with the increase of density, the matter structure is expected to change from "droplet" to "rod", "slab", "tube", "bubble" then to uniform. The existence of such "pasta" phases, instead of the crystalline lattice of nuclei, would affect several important processes in the supernova explosion by modifying the hydrodynamic properties and the neutrino opacity in supernova matter. It is also expected to influence the glitch of neutron stars via the change of the equation of state of the crust matter. Our first aim then is to study the nuclear "pasta" structure by means of a mean-field model which includes the Coulomb interaction in a fully consistent way.

At higher densities where kaon condensation may occur, it has been suggested that the structured mixed phase appears as a result of the first-order phase transition. If this is the case, we can expect the matter structure similar to the "pasta" phases [2]. In the first-order phase transitions with more than one chemical potential, the Maxwell construction (*coexisting separate phases* with local charge neutrality) does not necessarily fulfill the Gibbs conditions for the phase equilibrium,

$$T^{\mathrm{I}} = T^{\mathrm{II}}, \quad P^{\mathrm{I}} = P^{\mathrm{II}}, \quad \mu_B^{\mathrm{I}} = \mu_B^{\mathrm{II}}, \quad \mu_e^{\mathrm{I}} = \mu_e^{\mathrm{II}}; \tag{1}$$

the electron (charge) chemical potential takes different values between two phases, $\mu_e^{\rm I} \neq \mu_e^{\rm II}$, but we shall see that it means nothing but the difference in the electron number between two phases. When we naively apply the Gibbs conditions to these cases, we expect the *structured mixed phase* in a wide density range, where charge density as well as baryon density are no more uniform [1].

However, it has been suggested in recent papers that the Maxwell construction may still have a physical meaning by taking the hadron-quark matter transition as an example: the density region of the structured mixed phase is largely limited by the *Coulomb screening effect*, and results based on the Gibbs conditions become very close to the Maxwell construction curve [3]; note that if the Coulomb potential is properly included, it can give, combining with the charge chemical potential,

$$\rho_e = -(\mu_e - V_{\rm Coul})^3/3\pi^2 \tag{2}$$

for the electron charge density in a gauge invariant way. We see later that our calculation includes the Coulomb potential consistently with other equations of motion. The second aim of this paper is to clarify the Coulomb screening effect on the structure of matter in the first-order phase transitions.

## DENSITY FUNCTIONAL THEORY WITH RELATIVISTIC MEAN FIELD MODEL

We use density functional theory (DFT) with a relativistic mean field (RMF) model [4] in our study. The Coulomb potential is consistently included in the equations of motion. With this framework we can satisfy the Gibbs conditions in a proper way.

We start from the simple thermodynamic potential[2]:

$$\Omega = \Omega_B + \Omega_M + \Omega_K + \Omega_e, \tag{3}$$

$$\Omega_B = \int d^3r \left[ \sum_{i=p,n} \left( \frac{2}{(2\pi)^3} \int_0^{k_{Fi}} d^3k \sqrt{m_B^{*2} + k^2} - \rho_i \nu_i \right) \right], \tag{4}$$

$$\Omega_M = \int d^3r \left[ \frac{(\nabla\sigma)^2}{2} + \frac{m_\sigma^2 \sigma^2}{2} + U(\sigma) - \frac{(\nabla\omega_0)^2}{2} - \frac{m_\omega^2 \omega_0^2}{2} - \frac{(\nabla\rho_0)^2}{2} - \frac{m_\rho^2 \rho_0^2}{2} \right], \tag{5}$$

$$\Omega_K = \int d^3r \left[ -\frac{f_K^2 \theta^2}{2} \left[ -m_K^{*2} + (\mu_K - V_{\rm Coul} + g_{\omega K}\omega_0 + g_{\rho K}\rho_0)^2 \right] + \frac{f_K^2 (\nabla\theta)^2}{2} \right], \tag{6}$$

$$\Omega_e = \int d^3r \left[ -\frac{1}{8\pi e^2} (\nabla V_{\rm Coul})^2 - \frac{(V_{\rm Coul} - \mu_e)^4}{12\pi^2} \right], \tag{7}$$

where $\sigma, \omega_0, \rho_0$ are meson fields, $\mu_K = \mu_e$, $\nu_p = \mu_B - \mu_e + V_{\rm Coul} - g_{\omega N}\omega_0 - g_{\rho N}\rho_0$, $\nu_n = \mu_B - g_{\omega N}\omega_0 + g_{\rho N}\rho_0$, $m_B^* = m_B - g_{\sigma N}\sigma$, $m_K^* = m_K - g_{\sigma K}\sigma$, and the kaon field $K = f_K \theta/\sqrt{2}$ ($f_K$ : Kaon decay constant).[1] The parameters are chosen to reproduce the saturation properties of nuclear matter. From $\frac{\delta\Omega}{\delta\phi_i(\mathbf{r})} = 0$ ($\phi_i = \sigma, \rho_0, \omega_0, \theta$) or $\frac{\delta\Omega}{\delta\rho_i(\mathbf{r})} = 0$ ($i = n, p, e$), we get the equations of motion for fields as

$$-\nabla^2\sigma + m_\sigma^2\sigma = -\frac{dU}{d\sigma} + g_{\sigma N}(\rho_n^{(s)} + \rho_p^{(s)}) - 2g_{\sigma K} m_K f_K^2 \theta^2, \tag{8}$$

$$-\nabla^2\omega_0 + m_\omega^2\omega_0 = g_{\omega N}(\rho_p + \rho_n) + f_K^2 g_{\omega K}\theta^2(\mu_K - V_{\rm Coul} + g_{\omega K}\omega_0 + g_{\rho K}\rho_0), \tag{9}$$

$$-\nabla^2\rho_0 + m_\rho^2\rho_0 = g_{\rho N}(\rho_p - \rho_n) + f_K^2 g_{\rho K}\theta^2(\mu_K - V_{\rm Coul} + g_{\omega K}\omega_0 + g_{\rho K}\rho_0), \tag{10}$$

$$\nabla^2\theta = \left[{m_K^*}^2 - (\mu_K - V_{\rm Coul} + g_{\omega K}\omega_0 + g_{\rho K}\rho_0)^2\right]\theta, \tag{11}$$

$$\nabla^2 V_{\rm Coul} = 4\pi e^2 \rho_{\rm ch} \quad (\text{charge density } \rho_{\rm ch} = \rho_p + \rho_e + \rho_K), \tag{12}$$

$$\mu_p = \mu_B - \mu_e = \sqrt{k_{Fp}^2 + {m_B^*}^2} + g_{\omega N}\omega_0 + g_{\rho N}\rho_0 - V_{\rm Coul}, \tag{13}$$

$$\mu_n = \mu_B = \sqrt{k_{Fn}^2 + {m_B^*}^2} + g_{\omega N}\omega_0 - g_{\rho N}\rho_0. \tag{14}$$

---

[1] We here consider a linearized $KN$ Lagrangian for simplicity, which is not chiral-symmetric.

Note that the Poisson equation (12) is a highly nonlinear equation for $V_{\rm Coul}$, since $\rho_{\rm ch}$ in RHS includes it in a complicated way.

To solve the above coupled equations, we use the Wigner-Seitz cell approximation: the space is divided into equivalent cells with spherical shape (cylindrical (slab) shape in two (one) dimensional calculation). Each cell is charge-neutral and all the physical quantities in a cell are smoothly connected to those of the neighbor cell (zero gradient at the boundary). The cell is divided into grid points ($N_{\rm grid} \approx 100$) and the differential equations for fields are solved by a relaxation method with constraints of given baryon number and charge neutrality.

## PROPERTY OF FINITE NUCLEI

Before applying our model to nuclear matter, we check how it can describe finite nuclei. In this calculation, electron density is put to be zero and the boundary condition or the charge-neutrality condition is not imposed. However, the spherical-symmetry approximation is kept. In Fig. 1 (left panel) we show the density profiles of some typical nuclei. To get a better fit, we may need to include a surface term etc. Shell effects (see the drop at the center in $^{16}$O case) cannot be described by such a mean-field approach. By imposing the beta-equilibrium on the system, the most stable proton ratio can be obtained for a given mass number. Figure 1 (right panel) shows the mass-number dependence of the binding energy per nucleon and the proton-ratio. We can see that the bulk properties of finite systems (density, binding energy and proton ratio) are sufficiently reproduced.

Here we should note that we should adjust the sigma mass to be slightly smaller than that popularly used, i.e. 400 MeV to get such a good fit. If we use the popular value of $m_\sigma \approx 500$ MeV finite nuclei are overbound by about 3 MeV per nucleon. The sigma mass (or the omega mass) should be important for finite nuclei, i.e. non-uniform systems, since the meson mass is relevant to the interaction range and consequently affects, e.g., the nuclear surface tension.

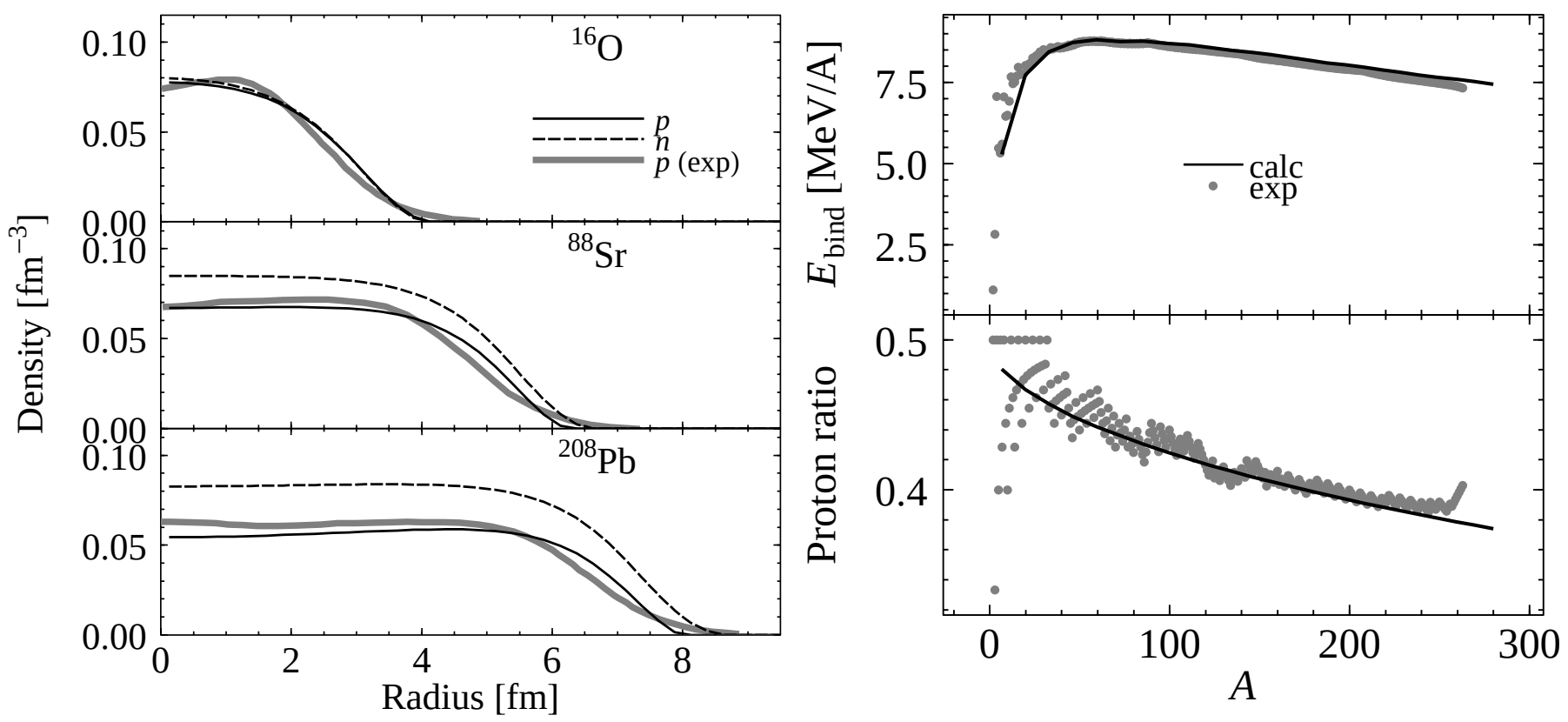

**FIGURE 1.** Left: the density profiles of typical nuclei. The proton densities (solid curves) are compared with the experiment. Right: binding energy per nucleon and proton ratio of finite nuclei.

## NUCLEAR "PASTA" AT SUB-NUCLEAR DENSITY

In the density region where nuclei are about to melt into uniform nuclear matter, it is expected that the energetically favorable mixed phase, which consists of a nucleon liquid and a nucleon gas, possesses interesting structures, such as rod-like and slab-like nuclei and rod-like and spherical bubbles, etc. These exotic structures are referred to as nuclear "pastas". The existence of the "pasta" phases instead of the crystalline lattice of nuclei would affect the supernova explosion or glitch phenomenon of neutron stars. Due to these importance and the curiosity, the "pasta" structure has been studied by several models. It is widely accepted that the appearance of the "pasta" structure is due to the balance of the Coulomb energy and the surface tension. However, the electron density has been always treated as an uniform background in the usual treatments. Here we study the nuclear "pasta" structure with our model which consistently treats the Coulomb potential and the electron distribution. Particularly we focus on symmetric nuclear matter (relevant to the supernova matter in the initial collapsing stage) where the electron density is comparable to the baryon density.

Figure 2 (left) shows some typical profiles of symmetric nuclear matter structure obtained with our model. The nuclear "pasta" is well described. One should note the non-uniform electron distribution. The phase diagram of matter structure is shown in Fig. 2 (middle). The size of the cell $R_{\rm cell}$ is optimized with precision of 1 fm, and the lowest energy solutions are chosen. We see, in the figure, that there appears no spherical hole configuration; this depends on the effective interaction used in the calculation.

To see the Coulomb screening effect, there are two possible ways: one is to solve equations of motion for fields neglecting the Coulomb potential $V_{\rm Coul}$ (afterward, the Coulomb energy is added to the total energy), and the other is only to discard $V_{\rm Coul}$ in RHS of the Poisson equation, consequently the electron distribution becomes uniform. The first one should be standard and very clear in its definition, while it is less meaningful in our model, where the matter structure is not assumed: without the Coulomb repulsion between protons the nuclear matter would always form a bulk droplet, independent of the cell size. In the second way, on the other hand, protons interact with each other and may form non-uniform structure through the balance of the nuclear surface tension and the Coulomb interaction in a uniform electron background. In the next section, however, we shall see the first way is more suitable since the electron field and the kaon field are treated on an equal footing and uniform kaon distribution is meaningless.

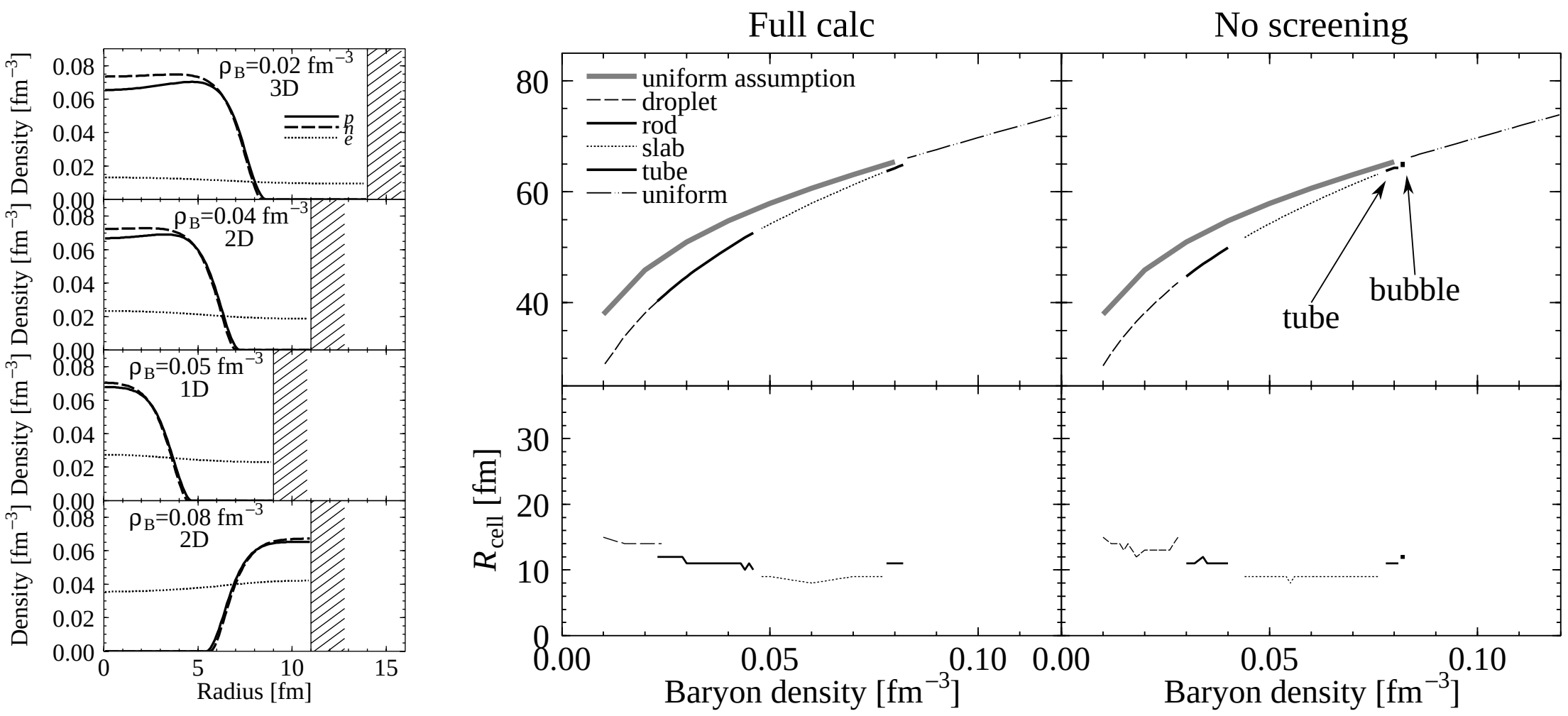

**FIGURE 2.** Left: examples of density profiles (droplet, rod, slab, and tube). Middle: binding energy per nucleon and the cell size of symmetric nuclear matter. Right: same as middle with uniform electron distribution.

We show in the right panel of Fig. 2 the results without the Coulomb screening (the second way), i.e. calculation with uniform electron background. The region of each structure (droplet, rod, etc.) is different from that of full calculation. Especially the "bubble" (spherical hole) appears in this case. However, such appearance of structure and its region is dependent on the very subtle energy difference, consequently on the effective interaction. The Coulomb screening effect on the bulk EOS (energy per baryon) is not so large.

## KAON CONDENSATION IN HIGH-DENSITY MATTER

Next we explore the high-density nuclear matter in beta-equilibrium. This matter corresponds to the inner core of a neutron star. If the Glendenning's claim is correct, the structured mixed phase develops in a wide density range from well below to well above the critical density of the first-order kaon condensation. Then nuclear matter should exhibit the similar structure change to the nuclear "pasta" phases: the kaonic droplet, the hole, and the uniform kaonic matter. In fact we observe such structures in our calculation (see Fig. 3). Note that the above result is only for three dimensional calculation; we considered only spherical configurations for the Wigner-Seitz cell. The "complex" configuration in the diagram means not a simple droplet or a hole structure but something like a shell shape or mixture of droplet and hole. So we may not expect such configuration to be realized, when two or one dimensional structure is taken into account.

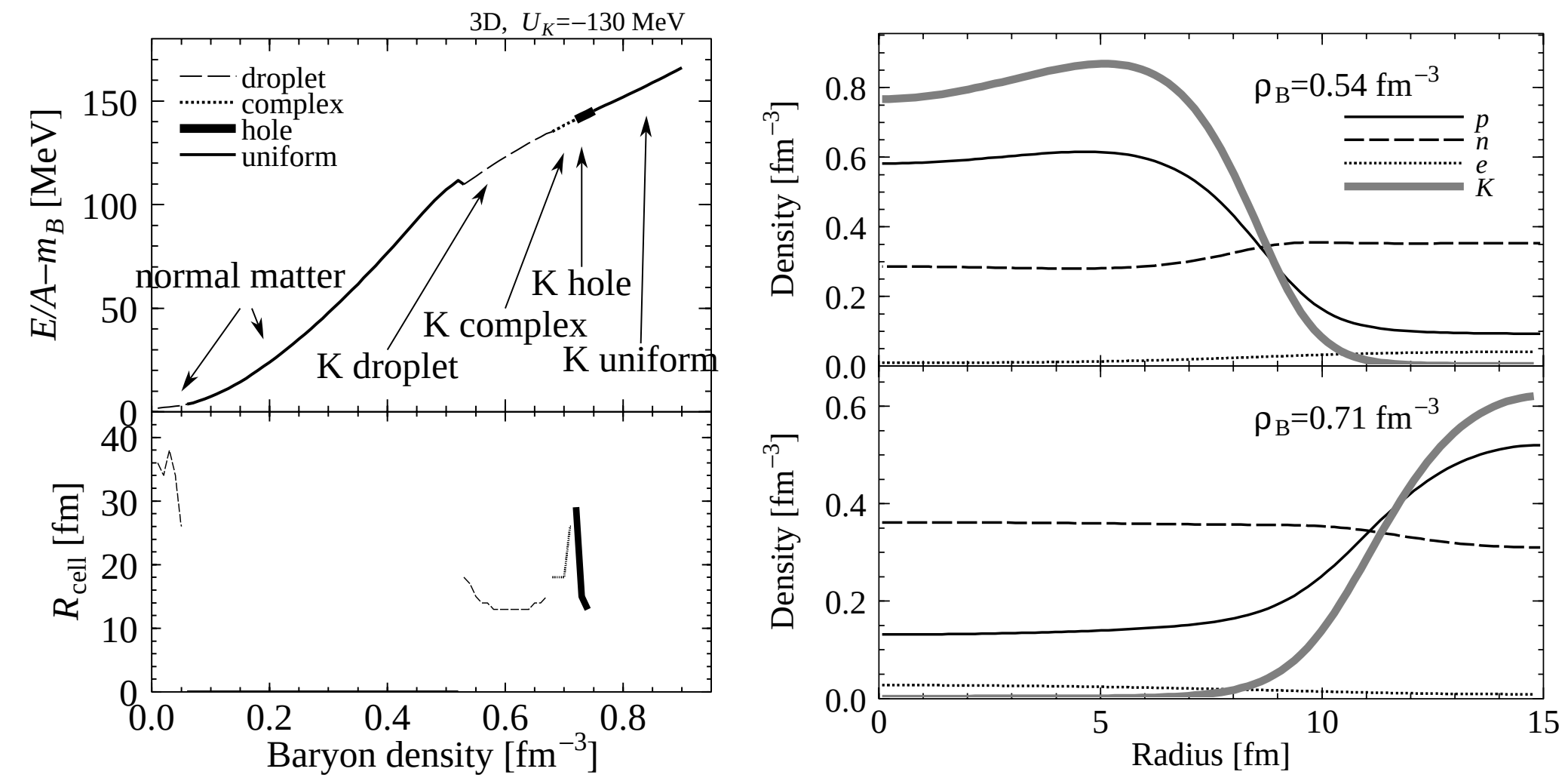

**FIGURE 3.** Left: binding energy per nucleon and the cell size of nuclear matter in beta equilibrium. Right: density profiles of kaonic matter. Droplet (upper panel) and hole (lower panel) configuration.

To demonstrate the Coulomb screening effects on the kaonic matter, we compare in Fig. 4 the phase diagrams in the $\mu_B$-$\mu_e$ plane with and without the Coulomb interaction. In this calculation the Coulomb potential $V_{\text{Coul}}$ is discarded in determining the density profile and the Coulomb energy calculated by this density profile is taken into account in the total energy. In Ref. [5], two cases, the Gibbs conditions and the Maxwell construction, are discussed. The case of the Gibbs conditions may lead to the structured mixed phase, while the Maxwell construction case to the phase separation of two bulk matters with local charge neutrality. Though we cannot definitely say now, the curve without the Coulomb interaction is similar to the one given by the Gibbs conditions and the curve with the Coulomb interaction to the one given by the Maxwell construction. If we look at the density profile, the local charge neutrality is more achieved in the case with the Coulomb interaction. These results suggest that the Maxwell construction is effectively meaningful due to the Coulomb screening.

## SUMMARY AND CONCLUDING REMARKS

We have discussed how nuclear matter structure changes during the first-order phase transitions. We took nuclear “pastas” and the structured mixed phase during the course of the kaon condensation as two examples.

Using a self-consistent framework based on DFT and RMF, we took into account the Coulomb interaction in a proper way. We have seen how the self-consistent inclusion of the Coulomb interaction changes the phase diagram. It becomes more remarkable in the case of the kaon condensation; the density range of the structured mixed phase is largely limited and thereby the phase diagram becomes similar to that given by the Maxwell construction. The density profiles there also suggest the phase separation of two bulk matter. On the other hand, it brings about rather little effect on the nuclear “pastas”. This is because the electron density is rather low and the Debye screening length is rather large compared with the size of the structure or the cell. Although the importance of such a treatment has been demonstrated for the quark-hadron matter transition[3], one of our new findings here is that we could figure out the peculiar role of the screening effect without introducing an "artificial" input for the surface tension; remember that we need to introduce a sharp boundary and its surface tension by hand in discussing the quark-hadron mixed phase. By using present results we can extract the surface tension numerically. Then we can discuss the present subjects again in a similar way to the previous studies, and may confirm them.

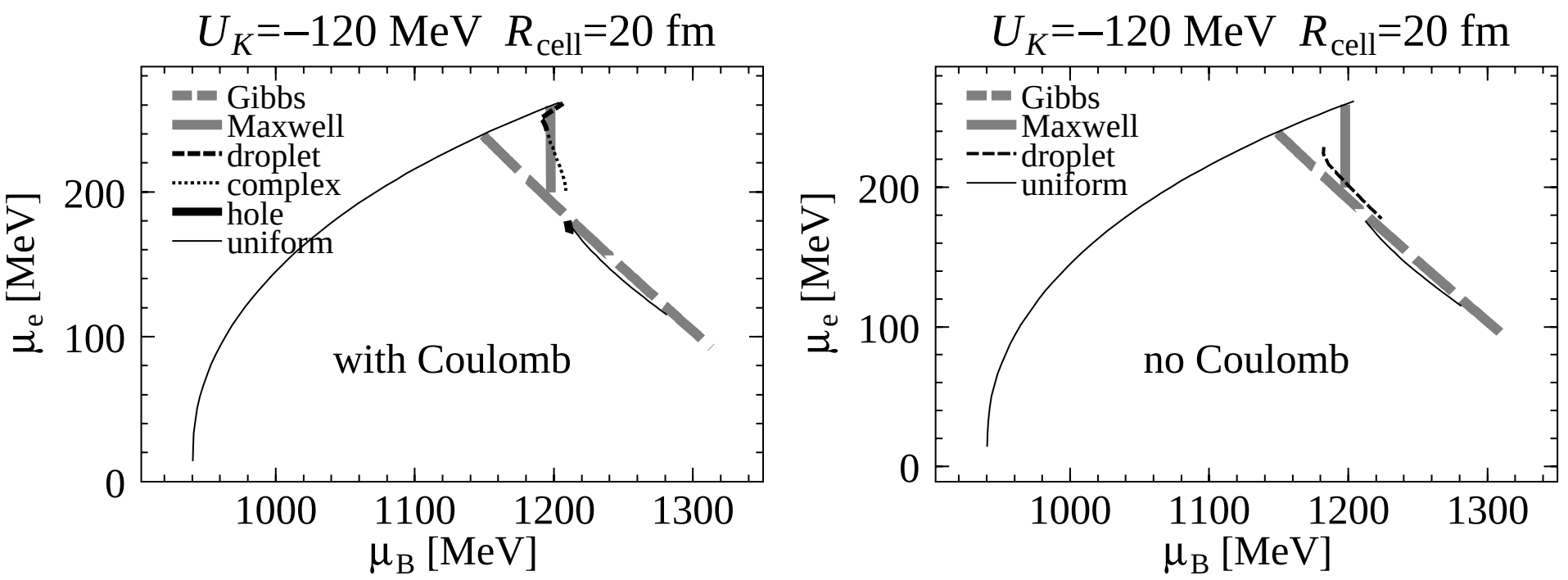

**FIGURE 4.** Phase diagram in the chemical potential plane. The cell radius is fixed to 20 fm. Left: full calculation. Right: Coulomb potential is discarded to determine the matter structure. Curves by Gibbs conditions and Maxwell construction are drawn in both panels. For simplicity both cases are calculated without the Coulomb interaction.

We have shown that our model can well reproduce the bulk properties of spherical nuclei. However, we should take into account the derivative terms for the densities to describe the surface region of the density profile in more realistic ways. This inclusion should be important not only quantitatively but also in the context of the structured mixed phase mentioned above.

We used a simple model to describe kaon condensation here. In return for it, we lost some interesting features related to chiral symmetry; actually it has been known that non-linearity of the kaon field causes a serious difficulty in satisfying the Gibbs conditions[6]. Then it would be interesting to see whether we have a consistent prescription without chiral models when the Coulomb interaction is properly taken into account.